\documentstyle[times]{ACMconf}
\def\lcb{\{}
\def\rcb{\}}
\title{Quantum Matching Theory ( with new complexity theoretic ,
combinatorial and topological insights on the nature of the Quantum Entanglement ) }
\author{Leonid Gurvits\\\
{\normalsize Los Alamos National Laboratory}\\
{\normalsize Los Alamos, NM  87545}\\
{\small email:  gurvits@lanl.gov}}
\begin{document}

%\date{}

\maketitle
 
%  THEOREM-LIKE ENVIRONMENTS
 
\newtheorem{THEOREM}{Theorem}[section]
\newenvironment{theorem}{\begin{THEOREM} \hspace{-.85em} {\bf :} 
}%
                        {\end{THEOREM}}
\newtheorem{LEMMA}[THEOREM]{Lemma}
\newenvironment{lemma}{\begin{LEMMA} \hspace{-.85em} {\bf :} }%
                      {\end{LEMMA}}
\newtheorem{FACT}[THEOREM]{Fact}
\newenvironment{fact}{\begin{FACT} \hspace{-.85em} {\bf :} }%
                      {\end{FACT}}
\newtheorem{COROLLARY}[THEOREM]{Corollary}
\newenvironment{corollary}{\begin{COROLLARY} \hspace{-.85em} {\bf 
:} }%
                          {\end{COROLLARY}}
\newtheorem{PROPOSITION}[THEOREM]{Proposition}
\newenvironment{proposition}{\begin{PROPOSITION} \hspace{-.85em} 
{\bf :} }%
                            {\end{PROPOSITION}}
\newtheorem{DEFINITION}[THEOREM]{Definition}
\newenvironment{definition}{\begin{DEFINITION} \hspace{-.85em} {\bf 
:} \rm}%
                            {\end{DEFINITION}}
\newtheorem{EXAMPLE}[THEOREM]{Example}
\newenvironment{example}{\begin{EXAMPLE} \hspace{-.85em} {\bf :} 
\rm}%
                            {\end{EXAMPLE}}
\newtheorem{CONJECTURE}[THEOREM]{Conjecture}
\newenvironment{conjecture}{\begin{CONJECTURE} \hspace{-.85em} 
{\bf :} \rm}%
                            {\end{CONJECTURE}}
\newtheorem{PROBLEM}[THEOREM]{Problem}
\newenvironment{problem}{\begin{PROBLEM} \hspace{-.85em} {\bf :} 
\rm}%
                            {\end{PROBLEM}}
\newtheorem{QUESTION}[THEOREM]{Question}
\newenvironment{question}{\begin{QUESTION} \hspace{-.85em} {\bf :} 
\rm}%
                            {\end{QUESTION}}
\newtheorem{REMARK}[THEOREM]{Remark}
\newenvironment{remark}{\begin{REMARK} \hspace{-.85em} {\bf :} 
\rm}%
                            {\end{REMARK}}
%\newenvironment{proof}{\noindent {\bf Proof:} \hspace{.677em}}%
%                      {}
 
%theorem
\newcommand{\thm}{\begin{theorem}}
%lemma
\newcommand{\lem}{\begin{lemma}}
%proposition
\newcommand{\pro}{\begin{proposition}}
%definition
\newcommand{\dfn}{\begin{definition}}
%remark
\newcommand{\rem}{\begin{remark}}
%example
\newcommand{\xam}{\begin{example}}
%conjecture
\newcommand{\cnj}{\begin{conjecture}}
%problem
\newcommand{\prb}{\begin{problem}}
%question
\newcommand{\que}{\begin{question}}
%corollary
\newcommand{\cor}{\begin{corollary}}
%proof
\newcommand{\prf}{\noindent{\bf Proof:} }
%end theorem
\newcommand{\ethm}{\end{theorem}}
%end lemma
\newcommand{\elem}{\end{lemma}}
%end proposition
\newcommand{\epro}{\end{proposition}}
%end definition
\newcommand{\edfn}{\bbox\end{definition}}
%end remark
\newcommand{\erem}{\bbox\end{remark}}
%end example
\newcommand{\exam}{\bbox\end{example}}
%end conjecture
\newcommand{\ecnj}{\bbox\end{conjecture}}
%end problem
\newcommand{\eprb}{\bbox\end{problem}}
%end question
\newcommand{\eque}{\bbox\end{question}}
%end corollary
\newcommand{\ecor}{\end{corollary}}
%end proof
\newcommand{\eprf}{\bbox}
%begin equation
\newcommand{\beqn}{\begin{equation}}
%end equation
\newcommand{\eeqn}{\end{equation}}
% white box
\newcommand{\wbox}{\mbox{$\sqcap$\llap{$\sqcup$}}}
%black box
\newcommand{\bbox}{\vrule height7pt width4pt depth1pt}
\newcommand{\qed}{\bbox}
% \sup will be used for superscript.
\def\sup{^}
% Defining Tchebyshef polynomial
\def\ra{\rightarrow}
\def\zok{\{0,1\}^k}
\def\zo{\{0,1\}}
\def\zom{\{0,1\}^m}
\def\zon{\{0,1\}^n}
\def\zor{\{0,1\}^r}
\def\zol{\{0,1\}^l}
\def\Tp{Tchebyshef polynomial}
\def\Tps{TchebysDeto be the maximafine $A(n,d)$ l size of a code with distance 
$d$hef polynomials}

%right arrow
\newcommand{\rarrow}{\rightarrow}
%left arrow
\newcommand{\larrow}{\leftarrow}
%right arrow

% R. Cleve definitions
%
\def\half{\textstyle{1 \over 2}}
\def\sqhalf{\textstyle{1 \over \sqrt{2}}}
\def\01{\{0,1\}}
\def\x{\times}
\def\ox{\otimes}
\def\xor{+}
\def\AND{}
\def\Q{\mbox{\sf Q}}
\def\e{\varepsilon}
\def\ket#1{\mbox{$| #1 \rangle$}}
\def\sqm#1{\textstyle{\sqrt{#1}}}
\def\half{\textstyle{1 \over 2}}
\def\one{\mbox{\boldmath $1$}}
\def\nl{\newline}
\def\ni{\noindent}
\def\ii{\hspace*{8mm}}
\def\ee{\vspace*{1mm}}
\def\eee{\vspace*{5mm}}
\def\F{{\cal F}}
\def\H{{\cal H}}
\def\loud#1{\noindent{\bf #1 }}
\def\EQ{\mbox{\it EQ\/}}
\def\EQprime{\mbox{\it EQ\/}^{\prime}}
\def\IP{\mbox{\it IP\/}}
\def\DISJ{\mbox{\it DISJ\/}}
\def\PARITY{\mbox{\it PARITY\/}}
\def\BAL{\mbox{\it BAL\/}}
\def\STAB{\mbox{\it STAB\/}}
\def\OR{\mbox{\it OR\/}}
\def\AND{\mbox{\it AND\/}}
\def\MAJ{\mbox{\it MAJORITY\/}}
\def\MODq{\mbox{\it MOD}_q}
\def\SIGMA{\mbox{\it SIGMA}}
\def\PI{\mbox{\it PI}}

\overfullrule=0pt
\def\setof#1{\lbrace #1 \rbrace}
\newcommand{\card}[1]{{\mathopen{|\!|}#1\mathclose{|\!|}}}
\newcommand{\pair}[1]{{\mathopen<#1\mathclose>}}
%\begin{titlepage}
%\maketitle
\section*{Abstract}
%\footnotesize
Classical matching theory can be defined in terms of matrices with nonnegative entries.
The notion of Positive operator , central in Quantum Theory , is a natural generalization
of matrices with nonnegative entries. Based on this point of view , we introduce a definition
of perfect Quantum (operator) matching . We show that the new notion inherits many "classical"
properties , but not all of them . This new notion goes somewhere beyound matroids .
For separable bipartite quantum states this new notion coinsides with the full rank
property of the intersection of two corresponding geometric matroids .
In the classical situation , permanents are naturally associated with perfects matchings.
We introduce an analog of permanents for positive operators, called Quantum Permanent and show how this generalization
of the permanent is related to the Quantum Entanglement. Besides many other things , Quantum Permanents
provide new rational inequalities necessary for the separability of bipartite quantum states .
Using Quantum Permanents , we give deterministic poly-time algorithm to solve
Hidden Matroids Intersection Problem and indicate some "classical" complexity difficulties
associated with the Quantum Entanglement.  
Finally , we prove that the weak membership problem for the convex set of separable
bipartite density matrices is NP-HARD.

\section{ Introduction and Main Definitions}
The (classical) Matching Theory is an important , well studied but still very active part
of the Graph Theory (Combinatorics) . 
The Quantum Entanglement is one of the central topics in Quantum Information Theory .
We quote from \cite{HSTT}   : "An understanding of entanglement seems to be at the heart of theories
of quantum computations and quantum cryptography , as it has been at the heart of quantum mechanics
itself . "
We will introduce in this paper a Quantum generalization of the Matching Theory and will
show that this generalization gives new and surprising insights on the nature of the Quantum Entanglement .
Of course , there already exist several "bipartite" generalizations of (classical)
bipartite matching theory . The most relevant to our paper is the Theory of Matroids ,
namely its part analyzing properties of intersections of two geometric matroids .
\dfn
Intersection of two geometric matroids
$MI(X,Y) = \{ (x_{i},y_{i}) , 1 \leq i \leq K \}$ is a finite family of distinct $2$-tuples of non-zero
$N$-dimensional complex vectors , i.e. $x_{i},y_{i} \in C^{N}$. \\
The rank of $MI(X,Y)$ is the largest integer $m$ such that there exist $1 \leq i_{1} <...<i_{m} \leq K $
with both sets $\{x_{i_{1}},...,x_{i_{m}} \}$ and $\{y_{i_{1}},...,y_{i_{m}} \}$ being
linearly independent. If  $Rank(MI(X,Y))$ is equal to $N$ then  $MI(X,Y)$ is called   matching .
The matroidal permanent $MP_{(X,Y)}$ is defined as follows :
\begin{eqnarray}
&MP_{(X,Y)} = : \nonumber   \\ 
&\sum_{1 \leq i_{1} < i_{2} <...< i_{N} \leq K}
 \det(\sum_{1 \leq k \leq N} x_{i_{k}}x_{i_{k}}^{\dagger}) 
 \det(\sum_{1 \leq k \leq N} y_{i_{k}}y_{i_{k}}^{\dagger})
\end{eqnarray}
\edfn
\rem
Let us denote linear space (over complex numbers ) of $N \times N$ complex matrices as $M(N)$ .
It is clear from this definition that $MI(X,Y)$ is matching iff $MP_{(X,Y)} > 0$. 
Moreover , $MI(X,Y)$ is matching iff the linear subspace $Lin(X,Y) \subset M(N)$ generated by the matrices
$\{x_{i}y_{i}^{\dagger} , 1 \leq i \leq K \}$ contains a nonsingular matrix  and , in general ,
 $Rank(MI(X,Y))$ is equal to the maximal matrix rank achieved in $Lin(X,Y)$ .
The following equality generalizes Barvinok's (\cite{Bar2} ) unbiased estimator for mixed discriminants :
\beqn
MP_{(X,Y)} = E (|\det(\sum_{1 \leq i \leq K} \xi_{i}x_{i}y_{i}^{\dagger})|^{2})
\eeqn
where $\{\xi_{i} ,1 \leq i \leq K \}$ are zero mean independent (or even $2N$-wise independent )
complex valued random variables such that $E(|\xi_{i}|^{2} = 1 , 1 \leq i \leq K $ .
It is not clear whether the analysis from \cite{Bar1} can be applied to $MP_{(X,Y)}$ .
\erem
\xam
Suppose that $ x_{i} \in \{e_{1},...,e_{N}\} , 1 \leq i \leq K $  , where $ \{e_{1},...,e_{N}\}$ is a standard basis in $C^{N}$.
Define the following positive semidefinite $N \times N$ matrices :
$$
Q_{i} = \sum_{ (e_{i}, y_{j}) \in (X,Y) } y_{j}y_{j}^{\dagger} , 1 \leq i \leq N .
$$
Then it is easy to see that in this case matroidal permanent coinsides with the mixed discriminant ,
i.e. $MP_{(X,Y)} = M(Q_{1} , \cdots , Q_{N} )$ where the mixed discriminant defined as follows :
\beqn 
M(Q_1,...Q_N) =   \frac{\partial^n}{\partial x_1
... \partial x_n} \det(x_1 Q_1 +....+x_N Q_N).
\eeqn
We will also use the following equivalent definition :
\beqn
M(Q_1,...Q_N) =  \sum_{\sigma,\tau \in S_N}
(-1)^{sign(\sigma\tau)} \prod_{i=1}^N Q_i(\sigma(i) , \tau(i)) ,
\eeqn
where $S_{n}$ is the symmetric group, i.e. the group of all permutations of the set $\{1,2, \cdots, N\}$. 
If matrices $Q_{i} , 1 \leq i \leq N $ are diagonal then their mixed discriminant is equal
to the corresponding permanent (\cite{GS}). 
\exam
Let us pose , before moving to Quantum generalizations , the following "classical" desision problem .
We will call it Hidden Matroids Intersection Problem ({\bf HMIP }) :
\prb Given linear subspace $ L \subset M(N) $  and a promise that $L$ has a ( hidden ) basis consisting of
rank one matrices. Is there exists poly-time deterministic algorithm to decide whether $L$ contains
a nonsingular matrix ? Or more generally , to compute maximum matrix rank achieved in $L$ ?
\eprb
Below in the paper we will assume that linear subspace $ L \subset M(N) $ in ({\bf HMIP }) is
given as a some rational basis in it. If this basis consists of rank one matrices then there is nothing
"hidden" and one can just apply standard poly-time deterministic algorithm computing
rank of intersection of two matroids. A natural (trivial) way to attack ({\bf HMIP }) would be
to exract a ( hidden ) basis consisting of rank one matrices. We are not aware about the complexity
of this extraction .  The following example shows that there exist linear subspaces
 $ L \subset M(N) $  having a rational real basis and a "rank one" basis but without rational
"rank one" basis :\\
Consider the following  $2 \times 2$   matrix 
$$
 A = \left( \begin{array}{cc}
		  0& -2 \\
		  0 & 1 \end{array} \right) ,		  
$$
and define linear subspace   $ IR \subset M(2) $  generated by  $A $ and the identity $I$ .\\
It is easy to see that \\
$Rank( aA + bI) \leq 1$  iff $a^{2} + 2b^{2}=0 $. \\
Therefore there are no rank one rational (complex) matrices in $IR$ . From the other hand rank one matrices
$$
C = \sqrt{2} I + i B ,  D = \sqrt{2} I -  i B
$$
form a  basis in $IR$ . \\
One of the main results of our paper is a positive answer to the nonsingularity part of ({\bf HMIP }) .
Moreover our algorithm is rather simple and does not require to work with algebraic numbers . \\
And , of course , we are aware about randomized poly-time  algorithms , based on Scwartz's lemma ,
 to solve this part of ({\bf HMIP }) . But for general linear subspaces , i.e. without extra promise ,
 poly-time deterministic algorithms are not known and the problem is believed to be "HARD" .
 To move to Quantum generalization , we need to recall several ,
standard in Quantum Information literature , notions .
\subsection{ Positive and completely positive operators ; bipartite density matrices and Quantum Entanglement}
\dfn
A positive semidefinite matrix $\rho_{A,B} : C^{N} \otimes C^{N} \rightarrow C^{N} \otimes C^{N} $
is called bipartite unnormalized density matrix \\
 ({\bf BUDM }) , if $tr(\rho_{A,B}) =1$ then this $\rho_{A,B}$ 
is called bipartite density matrix . \\
It is convinient to represent bipartite $\rho_{A,B} = \rho(i_{1},i_{2},j_{1},j_{2})$ as
the following block matrix :
\begin{equation} 
\rho_{A,B} = \left( \begin{array}{cccc}
		  A_{1,1} & A_{1,2} & \dots & A_{1,N}\\
		  A_{2,1} & A_{2,2} & \dots & A_{2,N}\\
		  \dots &\dots & \dots & \dots \\
		  A_{N,1} & A_{N,2} & \dots & A_{N,N}\end{array} \right) ,
\end{equation}
where $A_{i_{1},j_{1}} = : \{ \rho(i_{1},i_{2},j_{1},j_{2}) : 1 \leq i_{2},j_{2} \leq N \} ,
 1 \leq i_{1},j_{1} \leq N $ . \\
A ({\bf BUDM })   $\rho$ called {\bf separable } if
 \beqn
\rho = \rho_{(X,Y)} = : \sum_{1 \leq i \leq K} x_{i}x_{i}^{\dagger} \otimes y_{i}y_{i}^{\dagger} ,
\eeqn
and  {\bf entangled } otherwise . \\
If vectors $ x_{i}, y_{i} ; 1 \leq i \leq K $ in (6) are real then $\rho$ is called {\bf real separable } . \\
Quantum marginals defined as $\rho_{A} = \sum_{1 \leq i \leq N}A_{i,i} $ and  \\
$(\rho_{B}(i,j) = tr(A_{i,j} ) ;   1 \leq i , j \leq N) $ .\\

We will call ({\bf BUDM }) $\rho$ weakly separable
if there exists a separable $\rho'_{(X,Y)}$ with the same Image as  $\rho$ : $Im(\rho)= Im(\rho'_{(X,Y)})$. \\
A linear operator $T: M(N) \rightarrow M(N)$  called positive if $T(X) \succeq 0$ for all $X  \succeq 0$ ,
and strictly positive if $T(X) \succeq \alpha tr(X)I$ for all $X  \succeq 0$  and some $\alpha > 0$.
A positive operator T is called completely  positive if
\beqn 
T(X) = \sum_{1 \leq i \leq N^{2}} A_{i} X  A_{i}^{\dagger} ;     A_{i} ,  X  \in M(N)
\eeqn
Choi's representation of  linear operator $T: M(N) \rightarrow M(N)$  is a block matrix  $CH(T)_{i,j} = : T(e_{i}e_{j}^{\dagger}).$
Dual to $T$ respect to the inner product $<X,Y> =tr(XY^{\dagger}) $ is denoted as $T^{*}$.
Very usefull and easy Choi's result states that $T$ is completely  positive iff $CH(T)$ is   ({\bf BUDM }) .
Using this natural (linear) correspondence between completely  positive operators and ({\bf BUDM }) , we will
freely "transfer" properties of ({\bf BUDM }) to completely  positive operators . For example ,
a linear operator $T$ is called separable iff $CH(T)$ is separable , i.e.
\beqn
T(Z) = T_{(X,Y)}(Z) = \sum_{1 \leq i \leq K} x_{i}y_{i}^{\dagger}  Z  y_{i}x_{i}^{\dagger}
\eeqn
Notice that $CH(T_{(X,Y)})= \rho_{(X,Y)}$  and  $T_{(X,Y)}^{*}= T_{(Y,X)}$ .
\edfn
\rem
In light of definition (1.5) , we will represent linear subspaces $ L \subset M(N) \cong C^{N} \otimes C^{N}$ in ({\bf HMIP }) 
as images of weakly separable  ({\bf BUDM })  $\rho$ . And as the complexity measure we will use the number
of bits of (rational) entries of  $\rho$.
\erem
The next definition introduces the  quantum permanent  $ QP( \rho ) $ ,  the main tool to solve ({\bf HMIP }) .
Though it was not our original intention , it happens that $ QP( \rho _{(X,Y)}) = MP_{(X,Y)}$ . 
\dfn
We define quantum permanent,
$ QP( \rho ) $ , by the following equivalent formulas : 
\beqn
QP( \rho ) = :  \sum_{\sigma \in S_N} (-1)^{sign(\sigma)}M(A_{1,\sigma(1)},...,A_{N,\sigma(N)}) ;
\eeqn

\beqn
QP( \rho ) = \sum_{\tau_{1} , \tau_{2} , \tau_{3} \in S_N}(-1)^{sign(\tau_{1}\tau_{2}\tau_{3})}  \nonumber \\
\prod_{i=1}^N rho(i,\tau_{1}(i),\tau_{2}(i),\tau_{3}(i) )  ;
\eeqn

\begin{eqnarray}
QP( \rho ) & = & \frac{1}{N!}\sum_{\tau_{1} , \tau_{2} , \tau_{3} , \tau_{4} \in S_N}(-1)^{sign(\tau_{1}\tau_{2}\tau_{3})\tau_{4}} \nonumber \\
   & &\prod_{i=1}^N rho(\tau_{1}(i),\tau_{2}(i),\tau_{3}(i), \tau_{4}(i)) .  
\end{eqnarray}
\edfn

\rem
The representation (6) is not unique ,
it follows directly from the Caratheodory Theorem  that one always can choose $K \leq N^{4}$ in (6) .
Thus , the set of separable ({\bf BUDM }) , denoted by $Sep(N,N)$ ,  is a convex closed set .
As it is known that $Sep(N,N)$ has non-empty interiour , it follows from straigthforward dimensions counting
that for the "most" separable  ({\bf BUDM }) at least $K \geq \frac{N^{4}}{2N-1}$.
\erem
In the next proposition we summarize the properties of the quantum permanents we will need later in the paper .
\pro
\begin{enumerate}
\item
\beqn
QP( \rho_{(X,Y)} ) = MP_{(X,Y)} 
\eeqn

\item
\beqn
QP( \rho ) = <\rho^{\otimes N} Z , Z >  , 
\eeqn
where $\rho^{\otimes N}$ stands for a tensor product of
$N$ copies of $\rho$ , $< . , . >$ is a standard inner product and \\
$Z(j_{1}^{(1)},j_{2}^{(1)};...;j_{1}^{(N)},j_{2}^{(N)}) = \frac{1}{N!^{\frac{1}{2}}} (-1)^{sign(\tau_{1} \tau_{2})} $  \\
if $ j_{k}^{(i)} = \tau_{k}(i) ( 1 \leq i \leq N ) ;   \tau_{k} \in S_N  (k= 1,2) $ and zero otherwise . \\
( The equality (13) implies that if $ \rho_{1} \succeq  \rho_{2} \succeq 0$ then \\ 
$QP( \rho _{1}) \geq QP( \rho _{2}) \geq 0$  .)
\item
\beqn
 QP( (A_{1} \otimes A_{2} ) \rho (A_{3} \otimes A_{4} ) = \det(A_{1}A_{2}A_{3}A_{4}) QP(\rho)
\eeqn

\item
\beqn
QP( \rho_{A,B} ) = QP( \rho_{B,A} )
\eeqn

\end{enumerate}
\epro
\xam
Let us present a few cases when Quantum Permanents can be computed "exactly ".
They will also illustrate how universal is this new notion .
\begin{enumerate}
\item
Let $ \rho_{A,B} $ be a product state , i.e. $ \rho_{A,B} = C \otimes D$ .
Then $QP(C \otimes D) = Det(C) Det(D)$ .

\item 
Let $ \rho_{A,B} $ be a pure state , i.e. there exists a matrix  $( R=R(i,j) :  1 \leq i,j \leq N )$ \\
such that $\rho_{A,B}(i_{1},i_{2},j_{1},j_{2}) = R(i_{1},i_{2}) \overline{R(j_{1},j_{2})}$ .\\
In this case $QP( \rho_{A,B} ) = N! |Det(R)|^{2} $  . 

\item  
Define blocks of $ \rho_{A,B} $ as  $A_{i,j} = R(i,j) e_{i} e_{i}^{\dagger}$ . \\
Then $QP( \rho_{A,B} ) = Per(R) $ . 
\end{enumerate}
\exam
The next definition introduces Quantum Perfect Matching. 
\dfn
Let us consider a positive (linear) operator $T: M(N) \rightarrow M(N)$  ,
a map $G : C^{N} \rightarrow C^{N}$ ,
and the following three conditions :
\begin{enumerate}
\item  
$G(x) \in Im(T(xx^{\dagger}) $.
\item  
If $\{ x_1,...,x_N \}$ is a basis in $C^{N}$ then $\{ G(x_{1}),...,G(x_{N}) \}$ is also a basis,
i.e. the map $G$ preserves linear independence.
\item
If $\{ x_1,...,x_N \}$ is an orthogonal basis in $C^{N}$ then  \\
$\{ G(x_{1}),...,G(x_{N}) \}$ is  a basis .
\end{enumerate} .
We say that map $G$ is Quantum Perfect Matching for $T$ if it satisfies conditions (1,2) above ;
say map $G$ is Quantum Semi-Perfect Matching for $T$ if it satisfies conditions (1,3) above .

\edfn
In the rest of the paper we will address the following topics :
\begin{enumerate}
\item
 Characterization of Quantum Perfect Matchings in spirits of Hall's theorem .
\item
Topological and  algebraic properties of Quantum Perfect Matchings , i.e. properties of
maps $G$ in Definition (1.11).
\item 
Compelexity of checking whether given positive operator is matching .
\item 
Quantum (or Operator ) generalizations of Sinkhorn's iterations (in the spirit of \cite{lsw} , \cite{GY} , \cite{GS} ).
\item
van der Waerden Conjecture for Quantum Permanents.
\item
Connections between topics above and the Quantum Entanlement .
\item
Complexity to check the separability .
\end{enumerate}

\section{\bf Necessary and sufficient conditions for Quantum Perfect Matchings}
\dfn
A positive linear operator $T:M(N) \rightarrow M(N)$ called rank non-decreasing iff
\beqn
Rank(T(X)) \geq Rank (X) \mbox{ if} X \succeq 0 ;
\eeqn
and called indecomposable iff
\beqn
Rank(T(X)) > Rank(X) \mbox{ if} X \succeq 0 \mbox{ and} 1 \leq Rank(X) < N .
\eeqn
A positive linear operator $T:M(N) \rightarrow M(N)$ called doubly stochastic iff $T(I)=I$ and $T^{*}(I)=I$ ;
called $\epsilon$ - doubly stochastic iff $DS(T) =:  tr((T(I)-I)^{2}) + tr((T^{*}(I)-I)^{2}) \leq \epsilon^{2} $ .
\edfn
The next conjectures generalize Hall's theorem to Quantum Perfect Matchings .
\cnj
Assuming that the Axiom of Choice and the Continium Hypothesis hold,
a positive linear operator $T$ has Quantum Perfect Matching  iff it is rank non-decreasing .
\ecnj
 
\cnj
Assuming that the Axiom of Choice and the Continium Hypothesis hold,
a positive linear operator $T$ has Quantum Semi-Perfect Matching  iff it is rank non-decreasing .
\ecnj
\rem
We realize that the presence of the Axiom of Choice and the Continium Hypothesis in
linear finite dimensional result might look a bit weird . But 
we will illustrate below in this section that  for some completely positive entangled operators
corresponding Quantum semi-perfect matching maps $G$ are necessary quite complicated ,
for instance necessary discontinuos . Moreover Conjecture 1 is plain wrong , even for 
doubly stochastic indecomposable completely positive operators .
In separable and even weakly separable cases
one does not need "exotic axioms" and one can realize Quantum perfect matching map
it it exists as a linear nonsingular transformation through a rather simple use of Edmonds-Rado theorem .
\erem
The next Proposition(2.5) is a slight generalization of the corresponding result in \cite{lsw} .
\pro
Doubly stochastic operators are rank non-decreasing . If either $T(I)=I$ or $T^{*}(I)=I$   and   $DS(T) \leq N^{-1}$
then $T$ is rank non-decreasing . If  $DS(T) \leq (2N+1)^{-1}$  then $T$ is rank non-decreasing . 
\epro

\xam
Consider the following completely positive doubly stochastic operator $Sk_{3} : M(3) \rightarrow M(3)$ :
\beqn
Sk_{3}(X) = \frac{1}{2} A_{(1,2)}XA_{(1,2)}^{\dagger} + A_{(1,3)}XA_{(1,3)}{\dagger} + A_{(2,3)}XA_{(2,3)}{\dagger}
\eeqn
Here $\{ A_{(i,j)} , 1 \leq i < j \leq 3 \}$ is a standard basis in a linear subspace of $M(3)$
consisting of all skew-symmetric matrices , i.e. $A_{(i,j)} = : e_{i} e_{j}^{\dagger} - e_{i}e_{i}^{\dagger}$
and $\{ e_{i} , 1 \leq i \leq 3 \}$ is a standard orthonormal basis in $C^{3}$ .
It is easy to see that for a real normed $3$-dimensional column vector $x$ the image $ImSk_{3}(xx^{\dagger})$
is equal to the real orthogonal complement of $x$ , i.e. to the linear $2$-dimensional subspace $x^{\perp}$ 
of $R^{3}$
consisting of all real vectors orthogonal to $x$ . Suppose that $G$ is Quantum semi-perfect matching map ,
then $G(x) \in x^{\perp}$ and , at least , $G(x)$ is nonzero for nonzero vectors $x$.
By the well known topological result , impossibility to comb the unit sphere in $R^{3}$ , 
none of  Quantum semi- perfect matchings for $Sk_{3}$ is continuous.
It is not difficult to show that the operator $Sk_{3}$ is entangled .
A direct computation shows that 
\beqn
QP(CH(Sk_{3}) )= 0 
\eeqn
An easy "lifting" of this construction allows to get a similar example for all $N \geq 3$.
From the other hand , for $N=2$  all rank nondecreasing positive operators have linear
nonsingular Quantum perfect matchings . \\

\pro 
Assuming that the Axiom of Choice and the Continium Hypothesis hold, $Sk_{3}$ has a Quantum semi-perfect matching .
\epro
\prf (Sketch) Let us well order the projective  unit sphere $PS_2$ in $C^{3}$  : $ S_2  = (t_{\alpha} ;  \alpha \in \Gamma) $  
in such way that for any $\beta \in \Gamma$ the interval $(t_{\alpha} : \alpha \leq \beta)$ is at most countable .
Our goal is to build $(g_{\alpha} ;  \alpha \in \Gamma : g_{\alpha} \neq 0, g_{\alpha} \in t_{\alpha}^{\perp} )$  
such that if $(t_{\alpha_{1}},t_{\alpha_{2}},t_{\alpha_{3}})$
is orthogonal basis then $(g_{\alpha_{1}},g_{\alpha_{2}},g_{\alpha_{3}})$ is a basis . \\
As it usually happens in inductive consructions ,
we will inductively force an additional property :  $<g_{\alpha}, g_{\beta}> \neq 0 $   if $\alpha  > \beta$
and linear space $L(g_{\alpha}, g_{\beta})$  generated by $(g_{\alpha}, g_{\beta})$ is not equal to $L(t_{\alpha}, t_{\beta})$
if $<t_{\alpha}, t_{\beta}> = 0$.  In this , orthogonal case , $L(g_{\alpha}, g_{\beta}) = L(t_{\alpha}, t_{\beta})$
iff $g_{\alpha} = t_{\beta} \mbox{ and } g_{\beta} = t_{\alpha} $ . Using countability assumption , it is
easy to show that at each step of trasfinite  induction the set of 'bad" candidates has measure zero ,
which allows always to choose a "good" guy $g_{\gamma}$  without changing already constructed $(g_{\alpha} ;  \alpha < \gamma)$.
\eprf

The next Proposition shows that $Sk_{3}$ does not have Quantum perfect matchings !
\pro 
$Sk_{3}$ does not have Quantum perfect matchings 
\epro
\prf   Suppose that $G(.)$ is Quantum perfect matching for $Sk_{3}$ . We will get a contradiction
by showing that then there exists a basis $(b_{1},b_{2},b_{3})$ such that $<b_{1},b_{2}> = 0$
and $(G(b_{1}),G(b_{2}),G(b_{3}))$ are linearly dependent . For doing that , we need to
show that there exists an orthogonal basis $(O_{1},O_{2},O_{3})$ such that $O_{3}$
does not belong to $L(G(O_{1}),G(O_{2}))$. Indeed , if non-zero  \\
$d \in L(G(O_{1}),G(O_{2}))^{\perp}$ \\
then there is no basis $(G(O_{1}),G(O_{2}) , v) $ with  \\
$v \in  d^{\perp}=L(G(O_{1}),G(O_{2}))$   , but \\
$(O_{1},O_{2}, d$ is a basis since $<d,O_{3}>  \neq  0$ . \\
Take any non-zero $x$ and an orthogonal basis $\{y,z\}$ in $x^{\perp}$ such that \\
 $G(x) = (0,a_{1},a_{2})$  in
$\{x,y,z\}$ basis  and $a_{1} \neq 0 , a_{2} \neq 0$. \\
Let $G(y) = (b_{1},0,b_{2}) ,  G(z) = (c_{1},c_{2},0) $ . \\
Suppose that  $ z \in L(G(x),)G(y)), \mbox{ and } y \in L(G(x),)G(z)) $. \\
 Then $b_{1} = 0 \mbox{ and } c_{1} = 0$ .
This contradicts \\
to $((G(x),)G(y), G(z))$ being a basis . Thus there exists \\
 an orthogonal basis $(O_{1},O_{2},O_{3})$
 such that $O_{3}$ \\
 does not belong to $L(G(O_{1}),G(O_{2})) $ and we got a final contradiction.
\eprf
\exam

Next result shows that for weakly separable (and thus for separable) operators the situation
is very different.

\thm
Suppose that $T : M(N) \rightarrow M(N)$ is linear positive weakly separable operator ,
i.e. there exists a a family of rank one matrices $\{x_{1}y_{1}^{\dagger},...,x_{l}y_{l}^{\dagger} \} \subset M(N)$
such that for positive semidefinite matrices $X \succeq 0$ the following identity holds :
\beqn
     Im(T(X)) = Im ( \sum_{i=1}^l x_{i}y_{i}^{\dagger} X y_{i}x_{i}^{\dagger} )
\eeqn

Then the following conditions are equivalent :
\begin{enumerate}
\item $T$ is rank non-decreasing .
\item The rank of intersection of two geometric matroids $MI(X,Y)$ is equal to $N$.
\item The exists a nonsingular matrix $A$ such that $ Im(AXA^{\dagger}) \subset Im(T(X)) , X \succeq 0 $ .
\end{enumerate}
If , additionaly , $T$ is completely positive then these conditions are equivalent to existence
of nonsingular matrix $A$ such that operator $T'(X)= T(X) - AXA^{\dagger} $ is completely positive . \\
In this case $QP(CH(T)) \geq N! |Det(A)|^{2} > 0$ .
\ethm
\prf  Recall Edmonds-Rado Theorem for $MI(X,Y)$: \\
Rank of $MI(X,Y)$ is equal $N$ iff
\beqn
 \dim (L(x_{i}  ; i \in A ) + \dim (L(y_{j}  ; j \in \bar{A} ) \geq N ,
\eeqn
where $A \subset \{1,2,...,l\} $  and $ \bar{A}$ is a complement of $A$. \\
Suppose that rank of $MI(X,Y)$ is equal to $N$. Then 
$$ 
RankT(X) = \dim(L(x_{i}  ;  i \in A )) \mbox{ where }  A = : \{ i :   y_{i}^{\dagger} X y_{i} \neq 0 \}
$$
As $\dim (L(y_{j}  ; j \in \bar{A} ) \leq \ dim (Ker(X)) = N - Rank(X)$ hence , from Edmonds-Rado Theorem 
we get that $RankT(X) \geq N - (N - Rank(X)) = Rank(X) $ . \\
Suppose that $T$ is rank non-decreasing and for any $A \subset  \{1,2,...,l\} $ consider an orthogonal
proejctor $P \succeq 0$ on $L(y_{j}  ; j \in \bar{A} )^{\perp}$ . Then
$$
\dim(L(x_{i} :   i \in A ) )  \geq RankT(P) \geq Rank(P) = 
$$

$$
= N -\dim (L(y_{j}  ; j \in \bar{A} ) ).
$$
It follows from Edmonds-Rado Theorem that rank of  $MI(X,Y)$ is equal to $N$ . \\
All "equivalencies" follow now directly .
\eprf

\rem
Let us explain why Conjectures (1,2)  generalize Hall's theorem . Consider a square weighted incidence matrix $A_{\Gamma}$ of a
bipartite graph $\Gamma$ , i.e.$ A_{\Gamma}(i,j) > 0$  if  $i$ from the first part is adjacent to
$j$ from the second part and equal to zero otherwise. Then Hall's theorem can be immediately
reformulated as follows :
A perfect matching , which is just a permutation in this bipartite case , exists iff
$|A_{\Gamma} x|_{+} \geq |x|_{+}$ for any vector $x$ with nonnegative
entries , where $|x|_{+}$ stands for a number of positive entries of a vector $x$ .
One also can look at Theorem(2) as a Hall's like reformulation of Edmonds-Rado theorem .
\erem

\subsection{ A pleminary summary }
So far , we got neccessary and sufficient conditions for the existence of Quantum Perfect Matchings 
and presented , based on them , a new topological insight on the nature of the Quantum Entanglement.
It is not clear to us how crucial are "logical" assumptions in Prop.(2.7) . 
Theorem(2.9) shows that in separable (even weakly separable) case these assumptions are not needed .
The next question , which we study in the next sections , is about efficient , i.e.
polynomial time , deterministic algorithms to check the existence of Quantum Perfect Matchings .
We will describe  and analyse below in the paper 
a "direct" deterministic polynomial time algorithm
for weakly separable case . A complexity bound for a separable case is slightly better than
for just weakly separable case . Our algorithm is an operator generalization of
Sinkhorn's iterative scaling . We conjecture that without some kind of separability promise
checking the existence of Quantum Perfect Matchings is "HARD" even for completely positive operators.

\section { Operator Sinkhorn's iterative scaling }

Recall that for a square matrix $A = \{a_{ij}: 1 \leq i,j \leq N\}$ row
scaling is
defined as
$$
R(A) = \lcb \frac{a_{ij}}{\sum_j a_{ij}} \rcb \ , $$
column scaling as $C(A) = \lcb \frac{a_{ij}}{\sum_i a_{ij}} \rcb$
assuming that all
denominators are nonzero.

The iterative process $...CRCR(A)$ is called {\em Sinkhorn's iterative scaling} (SI).
There are two mainwell known properties of this iterative process , which we will generalize to positive
Operators.
\pro
\begin{enumerate}
\item Suppose that $A = \{a_{i,j} \geq 0: 1 \leq i, \ j \leq N\}$.  Then (SI)
convergess iff $A$ is matching, i.e., there exists a permutation $\pi$ such
that $a_{i,\pi (i)} > 0 \ (1 \leq i \leq N)$.
\item If $A$ is indecomposable, i.e., $A$ has a doubly-stochastic pattern
and is fully indecomposable in the usual sense , then (SI) converges exponentially
fast. Also in this case  there exist unique positive diagonal matrices $D_1 , D_2 , \det(D_2) = 1$ 
such that the matrix $D_1^{-1}AD_2^{-1}$ is doubly stochastic.
\end{enumerate}
\epro

\dfn [Operator scaling ]
Consider linear positive operator $T : M(N) \rightarrow M(N)$ . Define a new positive operator , Operator scaling , $S_{C_{1},C_{2}}(T) $ as :
\beqn
S_{C_{1},C_{2}}(T)(X) =: C_{1}T(C_{2}^{\dagger}XC_{2})C_{1}^{\dagger}
\eeqn
Assuming that both $T(I)$ and $T^{*}(I)$ are nonsingular we define analogs of row and column scalings :

\beqn
R(T) = S_{T(I)^{-\frac{1}{2}},I}(T) , C(T)= S_{I ,T^{*}(I)^{-\frac{1}{2}}}(T)
\eeqn
Operator Sinkhorn's iterative scaling (OSI) is the iterative process $...CRCR(T)$
\edfn
\rem Using Choi's representation of the operator $T$ in Definition(1.5) , we can define analogs of operator scaling (which are nothing but
so called local transformations ) and (OSI)  in terms of ({\bf BUDM }) : \\
\begin{eqnarray}
S_{C_{1},C_{2}}( \rho_{A,B}) =  C_{1} \otimes C_{2}( \rho_{A,B} ) C_{1} ^{\dagger} \otimes C_{2}^{\dagger}; \nonumber \\
R( \rho_{A,B}) = \rho_{A}^{-\frac{1}{2}} \otimes I  (\rho_{A,B} ) \rho_{A}^{-\frac{1}{2}} \otimes I , \nonumber \\
C( \rho_{A,B}) =      I   \otimes   \rho_{B}^{-\frac{1}{2}} ( \rho_{A,B} ) I   \otimes   \rho_{B}^{-\frac{1}{2}}.
\end{eqnarray}
\erem
Let us introduce a class of locally scalable functionals ({\bf LSF }) defined on a set of positive linear operators ,
i.e. functionals satisfying the following identity :
\beqn
\varphi(S_{C_{1},C_{2}}(T)) = Det(C_{1}C_{1}^{\dagger}) Det(C_{2}C_{2}^{\dagger}) \varphi(T)
\eeqn
We will call ({\bf LSF }) bounded if there exists a function $f$ such that $|\varphi(T)| \leq f(tr(T(I))$ .
It is clear that bounded ({\bf LSF }) are natural "potentials" for analyzing  (OSI) . Indeed , 
Let $T_{n} , T_{0} = T$ be a trajectory of (OSI) , $T$ is a positive linear operator . Then $T_{i}(I)=I$ for odd $i$
and $T_{2i}(I)^{*}=I  , i \geq 1$ . Thus if $\varphi(.)$ is ({\bf LSF })  then
\begin{eqnarray}
\varphi(T_{i+1}) = a(i) \varphi(T_{i}) ,  a(i) = Det(T_{i}^{*}(I))^{-1} \mbox{ if  } i \mbox{  is odd } , \nonumber \\
a(i) = Det(T_{i}(I))^{-1} \mbox{ if } i > 0 \mbox{  is even} .
\end{eqnarray}
As $tr(T_{i}(I)) = tr(T_{i}^{*}(I)) = N , i > 0$ , thus by the ariphmetic/geometric means inequality  we have that
$|\varphi(T_{i+1})|  \geq  |\varphi(T_{i})| $   and if $\varphi(.)$ is bounded  and $|\varphi(T)|  \neq 0$
then $DS(T_{n})$ converges to zero . \\

To prove a generalization of Statement 1 in Prop.(3.1) we need to "invent" a bounded ({\bf LSF })   $\varphi(.)$
such that  $\varphi(T) \neq 0$ iff operator $T$ is matching . We call such functionals responsible for matching .
It is easy to prove that $QP(CH(T))$ is a bounded ({\bf LSF }) . Thus if $QP(CH(T)) \neq 0$ then $DS(T_{n})$ converges to zero 
and , by Prop. (2.5) ,  $T$ is rank nondecreasing . From the other hand , $QP(CH(Sk_{3})) = 0$  and $Sk_{3}$ is
rank nondecreasing (even indecomposable ).  This is another "strangeness" of entangled operators ,
we wonder if it is possible to have "nice" ,  say polynomial with integer coefficients , responsible for matching ({\bf LSF }) ?
We introduce below responsible for matching  bounded ({\bf LSF }) and it is non-differentiable .
\dfn
For a positive operator $T: M(N) \rightarrow M(N)$, we define its capacity as  
\beqn
Cap(T) = \inf \{ Det(X): X \succ 0, Det(X) = 1\} \ . 
\eeqn
\edfn
It is easy to see that $Cap(T)$ is ({\bf LSF }) .  \\
Since $Cap(T) \leq Det(T(I) ) \leq (\frac{tr(T(I) )}{N})^{N}$ , \\
hence $Cap(T)$ is bounded ({\bf LSF }) . \\
\begin{lemma} A positive operator $T: M(N) \rightarrow M(N)$ is positive rank nondecreasing iff
$Cap(T) > 0$ .
\end{lemma}
\prf
Let us fix an orthonormal basis (unitary matrix) $ U =\{u_{1},...,u_{N}\} $ in $C^{N}$ 
and associate with positive operator $T$ the following positive operator :
\begin{equation}
T_{U}(X) = :\sum_{1 \leq i \leq N} T(u_{i}u_{i}^{\dagger}) tr(Xu_{i}u_{i}^{\dagger}) .
\end{equation}
(In physics words , $T_{U}$ is a decohorence respect to the basis $U$ , i.e.
in this basis applying $T_{U}$ to matrix $X$ is the same as applying $T$ to the
diagonal restriction of $X$.  )\\
It is easy to see that a positive operator $T$ is rank nondecreasing iff
operators $T_{U}$ are rank nondecreasing for all unitary $U$ . \\
And for fixed $U$ all properties of $T_{U}$ are defined by the following $N$-tuple
of $N \times N$ positive semidefinite matrices :
\begin{equation}
{\bf A}_{T,U} = : (T(u_{1}u_{1}^{\dagger}) , ..., T(u_{N}u_{N}^{\dagger}) .
\end{equation}
Importantly for us ,  $T_{U}$ is rank nondecreasing iff the mixed discriminant
$M(T(u_{1}u_{1}^{\dagger}) , ..., T(u_{N}u_{N}^{\dagger}) ) > 0 $. \\
Define capacity of ${\bf A}_{T,U}$ ,  
\begin{eqnarray*}
& Cap({\bf A}_{T,U}) = : \\
& \inf \{Det(\sum_{1 \leq i \leq N} T(u_{i}u_{i}^{\dagger}) \gamma_{i}) : \gamma_{i} > 0 , \prod_{1 \leq i \leq N}\gamma_{i} = 1 \} .
\end{eqnarray*}
It is clear from the definitions that    $Cap(T) $ is equal to infimum of $Cap({\bf A}_{T,U})$ over all unitary $U$. \\
One of the main results of \cite{GS} states that
\begin{eqnarray}
M({\bf A}_{T,U}) &=: &M(T(u_{1}u_{1}^{\dagger}) , ..., T(u_{N}u_{N}^{\dagger}) ) \leq Cap({\bf A}_{T,U}) \leq \nonumber \\
& & \leq  \frac{N^{N}}{N!} M(T(u_{1}u_{1}^{\dagger}) , ..., T(u_{N}u_{N}^{\dagger}) ).
\end{eqnarray}
As the mixed discriminant is a continuous (analytic ) functional and the group $SU(N)$ of unitary matrices is compact ,
we get the next inequality :
\beqn
\min_{U \in SU(N) } M({\bf A}_{T,U}) \leq Cap(T) \leq \frac{N^{N}}{N!}\min_{U \in SU(N) }M({\bf A}_{T,U})
\eeqn
The last inequality proves that $Cap(T) > 0$ iff positive operator$T$ is rank nondecreasing.
\eprf

So , the capacity is a bounded ({\bf LSF }) responsible for matching , which proves the next theorem :
\thm 
\begin{enumerate}
\item
Let $T_{n} , T_{0} = T$ be a trajectory of (OSI) , $T$ is a positive linear operator . Then $DS(T_{n})$ converges to zero iff
$T$ is rank nondecreasing .
\item
Positive linear operator $T$ is rank nondecreasing iff for all $\epsilon > 0$ there exists $\epsilon$-doubly stochastic
operator scaling of $T$ .
\end{enumerate}
\ethm
The next theorem generalizes second part of Prop. (3.1) and  is proved on almost the same lines as Lemmas 24,25,26,27 in \cite{GS} .
\thm
\begin{enumerate}
\item  There exist nonsingular matrices $C_{1},C_{2}$    such that $S_{C_{1},C_{2}}(T)$ is doubly stochastic
iff the infimum in ( 26)  is achieved .  \\
Moreover ,  if $Cap(T)  = Det(T(C)) $ where $ C \succ 0,  Det(C)=1 $   \\
then $S_{T(C)^{\frac{-1}{2}}, C^{\frac{1}{2}}} (T)$  is doubly stochastic . \\
Positive operator $T$ is indecomposable iff the infimum in ( 27)  is achieved and unique .
\item  Doubly stochastic operator $T$ is indecomposable iff  \\
 $ tr(T(X))^{2} \leq a \ tr(X)^{2}$ for some $0 \leq a < 1$
and all traceless hermitian matrices $X$.
\item  If Positive operator $T$ is indecomposable then $DS(T_{n})$ converges to zero with the exponential rate ,
i.e. $DS(T_{n}) \leq K a^{n}$ for some $K$ and $0 \leq a < 1$ .
\end{enumerate}
 \ethm
\section{Lower and upper bounds on Quantum Permanents }
The next proposition follows fairly directly from the second part of Prop.(1.9) and Cauchy-Schwarz inequality
\pro
Suppose that $\rho_{A,B}$ is ({\bf BUDM }). Then
\begin{eqnarray}
\max_{\sigma \in S_N} |D(A_{1,\sigma(1)},...,A_{N,\sigma(N)})|  = \nonumber  \\
D(A_{1,1},...,A_{1,N})
\end{eqnarray}
\epro

\cor
If $\rho_{A,B}$ is ({\bf BUDM })  then
\beqn
QP(\rho_{A,B}) \leq N! D(A_{1,1},...,A_{1,N}) \leq N! Det(\rho_{A}) .
\eeqn
Permanental part of Example(1.10) shows that  $N!$ is exact constant in both parts of (32) .
\ecor
The next proposition follows from the Hadamard's inequality : \\
if $X \succ 0 $ is $N \times N$ matrix then 
$Det(X) \leq  \prod_{i=1}^{N} X(i,i) $.
\pro 
If $X \succ 0$ then the following inequality holds :
\begin{eqnarray}
Det( \sum_{i=1}^{K}  x_{i}y_{i}^{\dagger} X y_{i}x_{i}^{\dagger} ) \geq  \nonumber \\
 Det(X) MP_{(X,Y) }.
\end{eqnarray}
\epro
\cor
Suppose that separable ({\bf BUDM })  $\rho_{A,B}$  is Choi's representation of completely positive operator $T$ . \\
Then for all $X \succ 0$ the next inequality holds :
\beqn
Det(T(X)) \geq QP(\rho_{A,B}) Det(X)
\eeqn
Since $\rho_{A} = T(I)$ , hence   $QP(\rho_{A,B}) \leq Det(\rho_{A})$  in separable case .
\ecor

Call ({\bf BUDM })  $\rho_{A,B}$  doubly stochastic if it is Choi's representation of completely positive
 doubly stochastic operator $T$ .  I.e. ({\bf BUDM })  $\rho_{A,B}$ is doubly stochastic iff
$\rho_{A} = \rho_{B} = I$ . As we already explained , the set of  separable ({\bf BUDM }) is
convex and closed . Thus the set of doubly stochastic separable ({\bf BUDM }) , $DSEP(N,N)$ , is a convex compact .
Define 
$$
\beta(N) = \min_{\rho \in DSEP(N,N)}  QP(\rho) .
$$
Then it follows that $\beta(N) > 0 $ for all integers $N$ .  The next conjecture is , in a sense , a third
generation of the famous van der Waerden conjecture . First generation is a permanental conjecture
proved by Falikman and Egorychev (\cite{fal} , \cite{ego}) in 1980  and second generation is Mixed discriminants conjecture
posed by R.Bapat \cite{bapat} in 1989 and proved by the author in 1999 \cite{gur}. Mixed discriminants conjecture
corresponds to block-diagonal doubly stochastic ({\bf BUDM }) . Any good lower bound on $\beta(N)$
will provide similarly to \cite{GS}  deterministic poly-time approximations for Matroidal permanents 
and new sufficient conditions for the Quantum Entanglement.
\cnj
\beqn
\beta(N) = \frac{N!}{N^{N}} ?
\eeqn
It is true for $N = 2$ .
\ecnj

\section{Polynomial time deterministic algorithm for ({\bf HMIP } ) }
We introduced Hidden Matroids Intersection Problem ({\bf HMIP }) as a well posed computer science
problem , which , seemingly , requires no "Quantum" background .
Also , we explained that ({\bf HMIP }) can be formulated in terms of weakly separable ({\bf BUDM }) .
Let us consider the following three properties of ({\bf BUDM })  $\rho_{A,B}$ .
( We will view this $\rho_{A,B} $  as Choi's representation of completely positive 
operator $T$ , i.e.      $\rho_{A,B} =   CH(T)$ . )
\begin{description}
\item[P1] 
$Im(\rho_{A,B})$  contains a nonsingular matrix . 
\item [P2]
The Quantum permanent $QP(\rho_{A,B}) > 0$ . 
\item [P3] 
Operator $T$ is rank nondecreasing .
\end{description}
We proved already that $P1 \longrightarrow P2 \longrightarrow P3$ and illustrated that
that the implication $P2 \longrightarrow P3$ is strict . In fact the implication $P1 \longrightarrow P2$  is also strict.
But , our Theorem (2.9), which is just an easy adoptation of Edmonds-Rado theorem , shows
that for weakly separable ({\bf BUDM })  the three properties $P1 , P2 , P3$ are equivalent .
Recall that to check $P1$ without the weak separability promise is the same as to check
whether given linear subspace of $M(N)$ contains a  nonsingular matrix  and it is very unlikely
that this desision problem can be solved in Polynomial Deterministic time .
Next , we will desribe and analyze Polynomial time deterministic algorithm to check whether $P3$ holds
provided that it is promised that $\rho_{A,B}$ is weakly separable . \\
In terms of Operator Sinkhorn's iterative scaling (OSI) we need to check if there exists $n$ such
that $DS(T_{n}) \leq \frac{1}{N} $ . If $ L = : \min\{n:  DS(T_{n}) \leq \frac{1}{N} \} $ is bounded
by a polynomial in $N$ and number of bits of $\rho_{A,B}$  then we have a Polynomial time Deterministic
algorithm to solve ({\bf HMIP } ) .
Algorithms of this kind for "classical" matching problem appeared independently in \cite{lsw} and \cite{GY} .
 In the "classical" case they are just another , conseptually simple , but far from optimal ,  poly-time algorithms to check whether
a perfect matching exists .  But for ({\bf HMIP } ) , our , Operator Sinkhorn's iterative scaling based approach
seems to be the only possibility ? \\
Assume that , without loss of generality , that all entries of $\rho_{A,B}$ are integer numbers and their
maximum magnitude is $Q$. Then $Det(\rho_{A}) \leq (QN)^{N}$ by the Hadamard's inequality .
If $QP(\rho_{A,B}) > 0$ then  necessary $QP(\rho_{A,B}) \geq 1$ for it is an integer number.
Thus 
$$
QP(CH(T_{1})) = \frac{QP(CH(T))}{Det(\rho_{A})} \geq (QN)^{-N}.
$$

Each $nth$ iteration ($n \leq L$ ) after the first one will multiply the Quantum permanent by $Det(X)^{-1}$ , 
where $X \succ 0 , tr(X)=N$ and $tr((X-I)^{2}) > \frac{1}{N} $ . Using results from \cite{lsw} ,
$Det(X)^{-1} \geq (1 -  \frac{1}{3N})^{-1} = : \delta $ . Putting all this together ,
we get the following upper bound on $L$ , the number of steps in (OSI) to reach the "boundary"
$DS(T_{n}) \leq \frac{1}{N}$  :
\beqn
\delta^{L}  \leq \frac{QP(CH(T_{L}))}{(QN)^{-N}}
\eeqn
It follows frm Prop.(4.2) and Cor.(4.4) that in weakly separable case $QP(CH(T_{L})) \leq N!$ \\
and in separable case $QP(CH(T_{L})) \leq 1$ .\\
Taking logarithms we get that in weakly separable case
\beqn
L \leq \approx  3N(N \ln(N) +N(\ln(N) + \ln(Q)) ;
\eeqn
and in separable case
\beqn
L \leq \approx  3N(N(\ln(N) + \ln(Q)) .
\eeqn
In any case , $L$ is polynomial in the dimension  $N$ and the number of bits $\log(Q)$. \\
To finish our analysis , we need to evaluate a complexity of each step of (OSI) . \\
Recall that $T_{n}(X) = L_{n}(T(R_{n}^{\dagger}XR_{n}))L_{n}^{\dagger}$ , \\
$T_{n}(I) =  L_{n}(T(R_{n}^{\dagger}R_{n}))L_{n}^{\dagger}$ and 
$T_{n}^{*}(I) =  R_{n}(T^{*}(L_{n}^{\dagger}L_{n}))R_{n}^{\dagger}$ . \\
To evaluate $DS(T_{n})$ we need to compute $tr((T_{n}^{*}(I)-I)^{2})$ for odd $n$ , \\
and $tr((T_{n}(I)-I)^{2})$ for even $n$ . \\
Define $P_{n}=L_{n}^{\dagger}L_{n} ,  Q_{n} = R_{n}^{\dagger}R_{n} $ .
It is easy to see that the matrix $T_{n}(I)$ is similar to $P_{n}T(Q_{n})$ ,
and $T_{n}^{*}(I)$ is similar to $Q_{n}T^{*}(P_{n})$ .  \\
As traces of similar matrices are equal , therefore to evaluate $DS(T_{n})$
it is sufficient to compute matrices $P_{n}, Q_{n} $. \\
But , $ P_{n+1} = (T(Q_{n}))^{-1} $   and  $ Q_{n+1} = (T^{*}(P_{n}))^{-1} $. \\
And this leads to  standard  , rational , matrix operations with $O(N^{3})$ per one
iteration in (OSI) . \\
Notice that our original definition of (OSI) requires computation of an operator square root .
It can be replaced by the Cholesky factorization , which still requires computing scalar
square roots . But our final algorithm is rational !
\section{Weak Membership Problem for a convex compact set of normalized
bipartite separable density matrices is NP-HARD }
One of the main research activities in Quantum Information Theory is a search for "operational"
criterium for the separability .  We will show in this section that , in a sense defined below , the problem is
NP-HARD even for bipartite normalized density matrices provided that each part is large (each "particle"
has large number of levels). First , we need to recall some basic notions from computational convex geometry.
\subsection{Algorithmic aspects of convex sets }
We will follow \cite{gr:lo:sc}.
\dfn
A proper ( i.e. with nonempty interior ) convex set $K \subset R^{n}$ called well-bounded $a$-centered
if there exist rational vector $a \in K$  and positive  (rational ) numbers $r,R$ such that
$B(a,r) \subset K$  and $K \subset B(a,R)$    (here $B(a,r) = \{x : \|x-a\| \leq r \}$ and $\| . \|$ is a standard eucleadian
norm in $R^{n}$ ) . Encoding length of such convex set $K$  is
$$
<K> = n + <r> + <R> + <a>  ,
$$
where  $<r> , <R> , <a>$  are the number of bits of corresponding rational numbers and rational vector .\\
Following \cite{gr:lo:sc} we define $S(K,\delta)$ as a union of all $\delta$-balls with centers belonging to $K$ ;
and $S(K,-\delta) = \{ x \in K :  B(x,\delta) \subset K \} $ . \\
\edfn
\dfn 
The Weak Membership Problem ($WMEM(K,y,\delta)$) is defined as follows : \\
Given a rational vector $y \in R^{n}$ and a rational number $\delta > 0$ either \\
(i) assert that $y \in S(K,\delta)$ , or \\
(ii) assert that $y \not\in S(K,-\delta)$ . \\

The Weak Validity Problem ($WVAL(K,c, \gamma , \delta ) $) is defined as follows : \\
Given a rational vector $y \in R^{n}$ , rational number $\gamma$ and a rational number $\delta > 0$ either \\
(i) assert that $ <c,x> = : c^{T}x \leq \gamma + \delta $ for all $x \in S(K,-\delta)$ , or \\
(ii) assert that $ c^{T}x \geq \gamma - \delta $  for some $x \in S(K,\delta)$ . \\
\edfn
\rem
Define $M(K,c) = : \max_{x \in K} <c,x>$  . It is easy to see that
\begin{eqnarray*}
& M(K,c) \geq M(S(K,-\delta),c) \geq M(K,c) - \|c\| \delta \frac{R}{r} ;  \\
 & M(K,c) \leq M(S(K,\delta),c) \geq M(K,c) + \|c\| \delta
\end{eqnarray*}
\erem
Recall that seminal Yudin - Nemirovskii theorem (\cite{YN}, \cite{gr:lo:sc}) implies that if there exists a deterministic
algorithm solving $WMEM(K,y,\delta)$  in $Poly( <K> + <y> + <\delta>)$  steps  then there exists a deterministic
algorithm solving  $WVAL(K,c, \gamma , \delta ) $ in $Poly( <K> + <c> +  <\delta>  + <\gamma>)$ steps. \\
Let us denote as $NSEP(M,N)$ a compact convex set of separable density matrices $\rho_{A,B} : C^{M} \otimes C^{N} \rightarrow C^{M} \otimes C^{N} $ ,
$tr(\rho_{A,B})=1$ ,  $M \geq N$ .  Recall that 
$$
NSEP(M,N) = CO(\{xx^{\dagger} \otimes yy^{\dagger} :  x \in C^{M} ,y \in C^{N}; \|x\| =  \|y\| = 1 \}) ,
$$
where $CO(X)$ stands for a convex hull generated by a set $X$.\\
Our goal is to prove that Weak Membership Problem for $NSEP(M,N)$ is NP-HARD .
As we are going to use Yudin - Nemirovskii theorem , it is sufficient to prove that $WVAL(NSEP(M,N),c, \gamma , \delta ) $ is NP-HARD respect
to the complexity measure $(M + <c> +  <\delta>  + <\gamma>)$
and to show that $<NSEP(M,N)>$ is polynomial in $M$.
\subsection {Geometry of $<NSEP(M,N)>$  }
First ,  $<NSEP(M,N)>$ can be viewed is a proper convex subset of the hyperplane in $R^{N^{2}M^{2}}$  . The standard euclidean 
norm in $R^{N^{2}M^{2}}$ corresponds
to the Frobenius norm for density matrices , i.e. $\|\rho\|_{F} = tr (\rho \rho^{\dagger})$.  The matrix $\frac{1}{NM} I  \in NSEP(N,N)$
and $\|\frac{1}{N^{2}} I -  xx^{\dagger} \otimes yy^{\dagger}\|_{F}   \leq 1 $ for all norm one
vectors $x,y$. Thus $NSEP(M,N)$ is covered by the ball $B(\frac{1}{NM} I , 1)$ . Next we will show that 
$B(\frac{1}{N^{2}} I , \sqrt{\frac{1}{N}}) \subset NSEP(N,N)$ .
Recall that $ \rho \in SEP(N,N) $ iff $tr(CH(T) \rho ) \geq 0$ for all positive operators $T : M(N) \rightarrow M(N) $.
This rather straightforward result was first proved in \cite{wor} . Let $\rho = \{A_{i,j} : 1 \leq i,j \leq N \} $ be a block matrix as in (5).
For a linear operator $\Psi : M(N) \rightarrow M(N) $ define \\
$\rho^{\Psi}  = \{\Psi(A_{i,j}) : 1 \leq i,j \leq N \} $ . \\
The following proposition is an easy reformulation of the above Woronowicz's criterium .
\pro
$\rho \in SEP(N,N)$ if and only if $\rho^{\Psi} \succeq 0$ for all positive operators $\Psi$ such that $\Psi(I) = I$ .
\epro
 \lem Suppose that $\Psi : M(N) \rightarrow M(N) $  is linear positive operator and $\Psi(I) = I$ . \\
 Then $\| \Psi (A)\|_{F} \leq \sqrt{N} \|A\|_{F}.$
 \elem 
 \prf For $A \in M(N)$ denote $\|A\|$ the  operator norm induced by a standard euclidean norm in $C^{N}$ (i.e. $\|A\|$ is the largest
 singular value of $A$ .  Recall that $ \|A\|^{2} \leq  \|A\|_{F}^{2} \leq N  \|A\|^{2}$ .
 Let  $B$ be a hermitian $N \times N$  complex matrix ,   then $ \|B\| I \succeq B \succeq - \|B\| I $. Thus using positivity and linearity
 we get that $ \|B\| I \succeq \Psi(B) \succeq - \|B\| I $ . We conclude that  
 \beqn
 \|\Psi(B)\| \leq \|B\|  \mbox{ for hermitian } B .
 \eeqn
  (The last inequality
 is in fact true for all matrices $B$ ). \\
 Let us consider an arbitrary $A  \in M(N)$ and decompose it uniquely  as $A = H_{1} + i H_{2} $ where matrices $H_{1} ,H_{2}$
 are hermitian :  $2 H_{1} = A + A^{\dagger} , 2 H_{2} = -i(A - A^{\dagger}) $. It is easy to check that 
 $$
 \|A\|_{F}^{2} = \|H_{1}\|_{F}^{2} + \|H_{2}\|_{F}^{2} .
$$
Therefore 
$$
\|\Psi (A)\|_{F}^{2} =   \|\Psi(H_{1})\|_{F}^{2} + \|\Psi(H_{2})\|_{F}^{2}  \leq N ( \|\Psi(H_{1})\|^{2} + \|\Psi(H_{2})\|^{2} ) .
$$
By (40 )   , we get that 
$$ 
\|\Psi(H_{1})\|^{2} + \|\Psi(H_{2})\|^{2} \leq \|H_{1}\|^{2} + \|\H_{2}\|^{2} \leq \|H_{1}\|_{F}^{2} + \|H_{2}\|_{F}^{2} =  \|A\|_{F}^{2}.
$$
Putting all this together , we finally get that 
\beqn
\|\Psi (A)\|_{F} \leq \sqrt{N} \|A\|_{F}
\eeqn
\eprf

\thm
Let $\Delta$ be a block hermitian matrix as in (5) . If $\|\Delta\|_{F} \leq \sqrt{\frac{1}{N}}$ then the the block matrix $I + \Delta$ is
separable.
\ethm
\prf
Let us consider positive linear operator  $\Psi : M(N) \rightarrow M(N) $  satisfying  $\Psi(I) = I$ . \\
Then $(I + \Delta)^{\Psi} = I  +  \Delta^{\Psi}$. Applying inequality (41) to each block of $\Delta$ and summing all of them
we get that $\|\Delta^{\Psi}\|_{F} \leq 1$ . As the matrix $\Delta^{\Psi}$ is hermitian , we conclude that $(I + \Delta)^{\Psi} \succeq  0$.
It follows from Proposition(6.4) that $I + \Delta$ is
separable.
\eprf

Summarizing , we get that   
$$
B(\frac{1}{N^{2}} I , \frac{1}{\sqrt{N}N^{2}} ) \subset NSEP(N,N) \subset B(\frac{1}{N^{2}} I , 1)
$$
and conclude that $<NSEP(N,N)> \leq  Poly(N) $.  It is easy to get from the last inequality that $<NSEP(M,N)> \leq  Poly(\max(N,M) $
It is left to prove that $WVAL(NSEP(M,N),c, \gamma , \delta ) $ is NP-HARD respect
to the complexity measure $(M+ <c> +  <\delta>  + <\gamma>)$ .
\subsection{ Proof of Hardness }
Let us consider the following hermitian block matrix :
\begin{equation} 
C = \left( \begin{array}{cccc}
		  0 & A_{1} & \dots & A_{M-1}\\
		  A_{1} & 0 & \dots & 0\\
		  \dots &\dots & \dots & \dots \\
		  A_{M-1} & 0 & \dots & 0\end{array} \right) ,
\end{equation}
i.e. $(i,j)$ blocks are zero if either $i \neq 1$ or $j \neq 1$ and $(1,1)$ block is also zero ;   $A_{1} ,..., A_{M-1}$
are real  symmetric $N \times N$ matrices .
\pro
\begin{eqnarray*}
& \max_{\rho \in NSEP(M,N)} tr (C \rho) = \\
& \max_{y \in R^{N} , \|y\| = 1}    \sum_{1 \leq i \leq M-1} (y^{T}A_{i}y)^{2} .
\end{eqnarray*}
\epro
\prf 
First , by linearity and the fact that the set of extreme points 
\begin{eqnarray*}
& Ext(NSEP(M,N)) = \\
& \{ xx^{\dagger} \otimes yy^{\dagger} : x \in C^{M} ,y \in C^{N} ; \|x\| =  \|y\| = 1 \}
\end{eqnarray*}
we get that 
\begin{eqnarray*}
& \max_{\rho \in NSEP(N,N)} tr (C \rho)  = \\
 & \max_{ xx^{\dagger} \otimes yy^{\dagger} :  x \in C^{M} ,y \in C^{N}; \|x\| =  \|y\| = 1 }  tr(C (xx^{\dagger} \otimes yy^{\dagger})) .
\end{eqnarray*}
But $tr(C (yy^{\dagger} \otimes xx^{\dagger})) = tr (A(y) xx^{\dagger})$ , where real symmetric $M \times M$ matrix $A(y)$ is defined as follows :
$$
A(y) = \left( \begin{array}{cccc}
		  0 & a_{1} & \dots & a_{M-1}\\
		  a_{1} & 0 & \dots & 0\\
		  \dots &\dots & \dots & \dots \\
		  a_{M-1} & 0 & \dots & 0\end{array} \right)  ;    a_{i} = tr (A_{i}  yy^{\dagger}) , 1 \leq i \leq M-1 .
$$		  
Thus 
\begin{eqnarray*}		  
& \max_{\rho \in NSEP(N,N)} tr (C \rho) = \\
& \max_{ yy^{\dagger} \otimes xx^{\dagger} : x \in C^{M} ,y \in C^{N} ; \|x\| =  \|y\| = 1 } tr (C \rho)   = \\
& \max_{\|y\| = 1 }  \lambda_{max} A(y) .
\end{eqnarray*}
(Above $\lambda_{max} A(y)$ is a maximum eigenvalue of $A(y)$) \\
It is easy to see $A(y)$ has only two non-zero eigenvalues $( d , -d )$ , where $d= \sum_{1 \leq i \leq M-1} (tr(A_{i}  yy^{\dagger}))^{2} $ .\\
As $A_{i} , 1 \leq i \leq N-1 $ are real symmetric matrices we finally get that 
$$
\max_{\rho \in NSEP(M,N)} tr (C \rho) = \max_{y \in R^{N} , \|x\| = 1} \sum_{1 \leq i \leq N-1} (y^{T}A_{i}y)^{2} .
$$
\eprf

Proposition(6.7) and Remark(6.3) suggest that in order  to prove NP-HARDness of \\
 $WVAL(NSEP(M,N),c, \gamma , \delta ) $  respect
to the complexity measure $M + <c> +  <\delta>  + <\gamma>$ it is sufficient to prove that the following problem of  is NP-HARD : \\
\dfn (RSDF problem)  Given  $k $  $l \times l$ real rational symmetric matrices $( A_{i} , 1 \leq i \leq l )$  and rational numbers $(\gamma , \delta)$ to check
whether
$$
\gamma + \delta \geq max_{x \in R^{l} , \|x\| = 1}f(x)  \geq \gamma - \delta ,  f(x) = :   \sum_{1 \leq i \leq l} (x^{T}A_{i}x)^{2} .
$$
respect to the complexity measure \\
$(l + k + \sum_{1 \leq i \leq l}<A_{i}>  + <\delta>  + <\gamma>)$ .
\edfn
It was shown in \cite{BN} that RSDF problem is NP-HARD provided $k \geq \frac{l(l-1)}{2} + 1$.
We summarize all this in the following theorem
\thm
The Weak Membership Problem for $NSEP(M,N)$ is NP-HARD if   $N \leq M \leq \frac{N(N-1)}{2} + 2$ .
\ethm 
\rem
It is easy exercise to prove that ({\bf BUDM }) $\rho_{A,B}$ written in block form (5) is real separable iff
it is separable and all the blocks in (5) are real symmetric matrices . It follows that , with obvious modifications ,
Theorem 6.9 is valid for the real separability too . \\
The construction (42) was inspired by Arkadi Nemirovski proof of NP-HARDness to check the positivity
of a given operator \cite{nem} .
\erem
\section{Concluding Remarks}
Many ideas of this paper were suggested by \cite{GS} .  The world of mathematical
interconnections is very unpredictable (and thus is so exciting) . The main technical result in a very recent breaktrough in
Communicational Complexity \cite{fos} is a rediscovery of particular , rank one , case of a general , matrix tuples scaling ,
 result proved in \cite{GS} with much simpler proof than in \cite{fos} .
Perhaps this our paper will produce something new in Quantum Communicational Complexity ? \\
We still don't know whether there is a deterministic poly-time algorithm to check whether given
completely positive operator is rank nondecreasing . And this question is related to lower bounds
on $Cap(T)$ provided that Choi's representation $CH(T) $ is an integer semidefinite matrix . \\
Theorem(6.9)  together with other
results from our paper gives a new , classical complexity based , insight on the nature
of the Quantum Entanglement and , in a sense , closes a long line of research in
Quantum Information Theory .   Still many open questions remained (for the author)  , for instance , is it still
NP-HARD for $(M,N)$ bipartite systems wnen $N$ is a fixed constant ?\\
We hope that the constructions introduced in this paper , especially Quantum Permanent ,
will have a promising future . The "third generation" of van der Waerden conjecture we introduced
above will require the "second generation" of Alexandrov-Fenchel inequalities \cite{Alexandrov}.
We think , that in general , mixed discriminants and mixed volumes should be studied (used )
more enthusiastically in the Quantum context .
After all , they are noncommutative generalizations of the permanent .... \\
Most of all , we hope that a reader will be able to "factor" our lousy english and to see the subject .  \\
It is my great pleasure to thank myLANL  colleagues Manny Knill and Howard Barnum  . \\
Finally , I would like to thank Arkadi Nemirovski for many enlightening discussions .

\end{document}